\documentstyle[aps,multicol,epsf]{revtex}

\begin{document}
\draft
\title{Macroscopic traffic models from microscopic car-following models }
\author{H. K. Lee,$^{1}$ H.-W. Lee,$^{2}$ and D. Kim$^{1}$}
\address{
$^1$ School of Physics,
Seoul National University, Seoul 151-742, Korea \\
$^2$ School of Physics, Korea Institute for Advanced Study,
207-43 Cheongryangri-dong, Dongdaemun-gu, Seoul 130-012, Korea \\
}
\maketitle

\begin{abstract}
We present a method to derive macroscopic fluid-dynamic models 
from microscopic car-following models via a coarse-graining procedure. 
The method is first demonstrated for the optimal velocity model.
The derived macroscopic model consists of
a conservation equation and a momentum equation,
and the latter contains a relaxation term, an anticipation term,
and a diffusion term. 
Properties of the resulting macroscopic model
are compared with those of the optimal velocity model 
through numerical simulations,
and reasonable agreement is found although
there are deviations in the quantitative level.
The derivation is also extended to general car-following models. 
\end{abstract}

\begin{multicols}{2}
\narrowtext

\section{Introduction}
\label{introduction}
For more 50 years, traffic flow has been
a subject of intense research effort~\cite{Herman63SA}.
While earlier studies were mostly conducted by traffic engineers, 
in the last decade the traffic flow problem has received great attention from
the physics community as well, largely due to 
the seminal works~\cite{Nagel92JPI,Biham92PRA,Kerner93PRE} in the early 90s,
which demonstrated that traffic flow can be regarded as
a driven nonequilibrium system.
There are empirical indications of multiple dynamic phases 
in the traffic flow and dynamic phase 
transitions~\cite{Treiterer74ISTTF,Kerner96bPRE,%
Neubert99PRE,Treiber00PRE,Lee00PRE}.
Several theoretical explanations~\cite{Nagatani97JPSJ,Lee98PRL,Helbing98PRL,%
Mitarai99JPSJ,Tomer00PRL,Nelson00PRE} for the empirical results
were suggested.
Also physical phenomena such as self-organized criticality and 
hysteresis~\cite{Wolf96TGF} were revealed.

Numerous traffic models have been investigated
(see Refs.~\cite{Nagel99ARCP,Chowdhury00PR,Helbing00preprint}
for recent reviews)
in relation to empirical data, and 
considerable progress has been achieved toward an understanding of
various traffic phenomena observed empirically.
Depending on the mathematical formulation used,
traffic models may be categorized into one of the following types:
car-following models, particle-hopping models,
coupled-map lattice models, gas-kinetic models,
and fluid-dynamic models.
The first three types use a microscopic approach
while the last type uses a macroscopic one.
The approach used in the gas-kinetic models is intermediate
and may be called mesoscopic.

Recently it was suggested~\cite{Hermann98PhysicaA,Hayakawa98PTPS} that 
different types of traffic models may belong to
the same ``universality'' class in the sense
that they share qualitatively similar properties.
More recently, a nonlocal fluid-dynamic model was derived from
a gas-kinetic model~\cite{Treiber99PRE}.
These reports motivate further studies on
mutual relationship between different types of traffic models.

In this paper, we address the relationship between
microscopic car-following models and macroscopic 
fluid-dynamic models.
Specifically we use a coarse-graining procedure (Sec.~\ref{formulation})
to derive a macroscopic model (Sec.~\ref{optimal})
from the microscopic optimal velocity model,
a particular case of the car-following-type model.
The resulting macroscopic model consists of 
the continuity equation [Eq.~(\ref{VehicleConservation})]
and a momentum equation [Eq.~(\ref{VelocityTotalDeriv_Final})].
The momentum equation contains
a relaxation term, a density gradient term, 
and a diffusion term,
similar to the fluid-dynamic model proposed in Ref.~\cite{Kerner93PRE}.
It is shown that both the density gradient term
and the diffusion term arise from a {\it directed} influence
due to the breakdown of the balanced action-reaction.
This is contrary to heuristic derivations~\cite{Kerner93PRE}, 
in which the density gradient term is attributed to the velocity variance.
It also provides an origin of the diffusion term
assumed in many fluid-dynamic models.
In Sec.~\ref{extension}, the derivation is extended to
general car-following-type models.
In Sec.~\ref{numerical}, 
the macroscopic model derived from the microscopic optimal velocity model
is examined numerically in comparison with
the optimal velocity model.
Section~\ref{discussion} concludes the paper.
Some technical details are presented in 
Appendixes~\ref{nonlinear}, \ref{linear_regime}, and \ref{differences}.

We remark that a different scheme 
to construct macroscopic models from microscopic car-following models
was proposed recently~\cite{Helbing01MCM}.
The macroscopic fields $\rho$ and $v$ are defined 
via an interpolation procedure instead of a coarse-graining procedure.
The resulting momentum equation is {\it nonlocal}, 
while our momentum equation is local.
Also $\rho$ and $v$ defined in such a way
do not strictly satisfy continuity equation~(\ref{VehicleConservation}),
while
the continuity equation is an exact identity 
in the coarse-graining-based scheme.

\section{General formulation}
\label{formulation}
In order to derive macroscopic traffic equations
from microscopic ones, 
we first introduce two microscopic field variables,
density field $\hat{\rho}(x,t)$ and flux field $\hat{q}(x,t)$,
\begin{eqnarray}
\hat{\rho}(x,t) & \equiv & \sum_n \delta \left (y_n(t)-x\right ) \ ,
  \nonumber \\
\hat{q}(x,t) & \equiv & \sum_n \dot{y}_n(t) \delta \left (y_n(t)-x\right ) \ ,
\end{eqnarray}
where $y_n(t)$ is the coordinate 
of the $n$th vehicle at time $t$  
with $y_1 < y_2 < \cdots < y_{n-1} < y_n < y_{n+1} <\cdots$\ . 
When traffic dynamics does not depend on 
third or higher order time derivatives of $y_n(t)$,
these two fields specify 
the status of traffic flow completely. 

A natural way to obtain macroscopic description
is to coarse grain these fields.
We introduce a coarse graining envelope function $\phi(x,t)$
which is non-negative valued, peaked at $(x,t) = (0,0)$, 
and normalized as $\int dx dt \phi(x,t) = 1$.
The coarse grained density $\rho(x,t)$ and flux $q(x,t)$ 
can be defined as
\begin{eqnarray}
   \rho(x,t) & \equiv & \int dx^\prime dt^\prime \, \phi(x-x^\prime,t-t^\prime)
   \hat{\rho}(x',t') \ ,
 \nonumber \\
 q(x,t) & \equiv & \int dx' dt' \, \phi(x-x',t-t') \hat{q}(x',t') \ .
   \label{rhoandQ} 
\end{eqnarray}
These two coarse grained fields specify
the {\it macroscopic} status of traffic flow.

Next we derive equations that govern 
the time evolution of $\rho(x,t)$ and $q(x,t)$.
For the evolution of $\rho(x,t)$, one finds
\begin{equation}
   {\partial\over{\partial t}}\rho(x,t) + 
   {\partial\over{\partial x}} q(x,t) = 0 ,
   \label{VehicleConservation}
\end{equation}
which describes the local conservation of vehicles 
in the coarse-grained description.
This equation can be verified from Eq.~(\ref{rhoandQ}) 
using integration by parts and change of variables.

Derivation of the dynamic equation for $q(x,t)$ is less straightforward.
After some algebra, one obtains
\begin{equation}
   {{\partial }\over{\partial t}}q(x,t) = 
   \rho(x,t) \left < \ddot{y}_n(t^\prime) \right >_{(x,t)}
   - {{\partial}\over{\partial x}} \left [ \rho(x,t) \left <{\dot{y}_n}^2
   (t^\prime)
   \right >_{(x,t)} \right ],
   \label{roundQroundT} 
\end{equation}
where the bracketed average of a quantity $O_n(x',t')$
is defined as follows; 
\begin{eqnarray}
   \left < O_n(x^\prime,t^\prime) \right >_{(x,t)} &\equiv& 
   {1\over \rho(x,t)}
   \int dx^\prime dt^\prime \, \phi(x-x^\prime,t-t^\prime) \nonumber \\
   && ~~~ \times\sum_n O_n(x^\prime,t^\prime)
   \delta(y_n(t^\prime)-x^\prime).
   \label{AverageDefinition}
\end{eqnarray}
Note that $x'$, $t'$, and $n$ inside the brackets are dummy variables,
while the label $(x,t)$ in the subscript of the bracket notation
represents a spatiotemporal position where 
the average is evaluated. This label will be omitted
in the rest of the paper when its omission does not cause confusion.

Here it is useful to introduce another macroscopic field $v(x,t)$,
\begin{equation}
v(x,t)\equiv \left< \dot{y}_n(x',t') \right>=q(x,t)/\rho(x,t),
\label{velocity}
\end{equation}
which represents some kind of macroscopic velocity, 
whose precise meaning depends on $\phi(x,t)$.
Two particular coarse graining schemes are good for illustration:
spatial coarse graining $\phi(x,t)=\delta(t)\Theta(X/2-|x|)/X$
and temporal coarse graining $\phi(x,t)=\delta(x)\Theta(T/2-|t|)/T$,
where $\Theta(x)$ is the step function which
is one for $x>0$ and zero for $x<0$.
For the spatial coarse graining, 
$v(x,t)$ becomes 
$$
v(x,t)={\sum'_n \dot{y}_n(t) \over \sum'_n 1}\, ,
$$
where the primed summation runs over the vehicles 
in the range $(x-X/2,x+X/2)$ at time $t$.
The denominator is equal to the total
number of vehicles within the range
and thus $v(x,t)$ represents the {\it arithmetic mean velocity}.
For the temporal coarse graining, on the other hand,
it can be verified that
$$
{1 \over v(x,t)}={\sum'_n \left[ \dot{y}_n(t_n(x))\right]^{-1} 
\over \sum'_n 1}\, ,
$$
where the primed summation now runs over the vehicles
that reach the point $x$ within the time
interval $(t-T/2,t+T/2)$, and 
$t_n(x)$ represents the time at which
the $n$th vehicle reaches the position $x$.
Here $\dot{y}_n(t)\ge 0$ is assumed.
Thus $v(x,t)$ represents the {\it harmonic mean velocity}
measured at local detectors.

It is straightforward to rewrite Eq.~(\ref{VehicleConservation}) 
in terms of $\rho$ and $v$ instead of $\rho$ and $q$.
Also expressing Eq.~(\ref{roundQroundT}) in terms of $\rho$ and $v$, 
one obtains
\begin{equation}
   \rho \left ( {{\partial v} \over {\partial t}} +
   v{{\partial v}\over{\partial x}} \right ) =
   \rho \left < \ddot{y}_n(t^\prime) \right >
   - {{\partial}\over{\partial x}} ( \rho \theta ),
   \label{VelocityTotalDerivative}
\end{equation}
where 
$$
\theta(x,t) \equiv \left<\dot{y}^2_n(t^\prime) \right> 
- v^2(x,t)
$$ 
measures the degree of microscopic velocity variation.
Note that the left-hand side of Eq.~(\ref{VelocityTotalDerivative})
corresponds to the total derivative
$Dv/Dt\equiv \partial v/\partial t+v\partial v/\partial x$.
Thus the two terms on the right-hand side can be interpreted as
macroscopic force densities.
The first term corresponds to the coarse-grained average of
microscopic ``forces'' that act on each vehicle. 
The second term, on the other hand, arises
from the coarse graining itself.
In equilibrium systems,
$\theta$ is proportional to the local temperature,
and the second term represents the force due to thermal gradient.

The remaining job is to express the force terms in terms of $\rho$ and $v$.
However, it is well known that a rigorous treatment of
the force terms generates an infinite sequence of dynamic equations.
Thus we instead develop approximations of the force terms in Sec.~\ref{optimal},
so that 
Eqs.~(\ref{VehicleConservation}) and (\ref{VelocityTotalDerivative}) form a closed set of equations.
This scheme is partly motivated by the absence of
empirical indication that the dynamics of the forces
is important.  

A procedure to derive a macroscopic model is illustrated for 
the optimal velocity model in Sec.~\ref{optimal}
and for general car-following models in Sec.~\ref{extension}.
In both sections, traffic states are assumed to
be almost homogeneous.
In this linear regime, products of differentiated quantities such as 
$\prod_{m=1}^{M}(\partial^{l_m} O_m/ \partial x^{l_m})$
become progressively smaller as $M$ increases,
where $l_m$ are integers and $O_m$ are
arbitrary functions of $\rho$ and $v$.
Therefore, it is sufficient to retain terms 
with $M=0$ or $1$ only, which simplifies
the construction of a macroscopic description considerably.
In this sense, terms with $M=0$ or $1$ can be called {\it linearly relevant}
terms, and terms with $M \geq 2$ {\it linearly irrelevant} terms.
Properties in the linear regime such as the dispersion relation
for small amplitude waves depend on linearly relevant terms only.
Effects of the linearly irrelevant terms with $M=2$ are discussed in
Appendix~\ref{nonlinear}.

\section{Optimal velocity model}
\label{optimal}
We first study the optimal velocity model~\cite{Bando95PRE}
\begin{equation}
   \ddot y_n(t) = \lambda \left[V_{\rm op}
   (\Delta y_n(t)) 
   - \dot y_n(t) \right],
   \label{OptimalVelocityModel}    
\end{equation}
where the constant $\lambda$ represents a driver's sensitivity
and $\Delta y_n\equiv y_{n+1}-y_n$ is
the coordinate difference between the vehicle $n$ and its preceding
vehicle $n+1$.
$V_{\rm op}(\Delta y)$ is the optimal velocity to which
drivers want to adjust their speed. 
An example is $V_{\rm op}(\Delta y)=\tanh(\Delta y-2)+\tanh 2$
used by Bando {\it et al}.~\cite{Bando95PRE}. 
Here we will assume neither a particular functional form for 
$V_{\rm op}(\Delta y)$
nor a particular value for $\lambda$~\cite{Komatsu95PRE}.
 
The coarse graining of Eq.~(\ref{OptimalVelocityModel}) leads to
\begin{equation}
\left<\ddot y_n\right>=
\lambda \left[\left<V_{\rm op}(\Delta y_n)\right>-v\right].
\label{averageForce}
\end{equation}
The expansion of $\left<V_{\rm op}(\Delta y_n)\right>$ with respect to 
$\left<\Delta y_n\right>$ gives
\begin{eqnarray}
   \left<V_{\rm op}\left(\Delta y_n\right)\right> 
   &=& \sum_{m=0}^{\infty}
   {1\over{m!}}V_{\rm op}^{(m)}\left(\left<\Delta y_n\right>\right)
   \left<(\Delta y_n-\left<\Delta y_n\right>)^m\right>
      \nonumber \\
   &\equiv& V_{\rm op}\left(\left<\Delta y_n\right>\right)
      + \sum_{m=2}^\infty I_m,
   \label{OptimalVelocityExpansion} 
\end{eqnarray}   
where $I_m$ is the term that is proportional to 
$\left < (\Delta y_n-\left<\Delta y_n\right>)^m \right >$. 
Here $I_1$ is absent
since $\left < (\Delta y_n-\left<\Delta y_n\right>)\right >=0$.
Note that the leading correction $I_2$ 
compensates for the difference $\langle V_{\rm op}(\Delta y_n) \rangle-
V_{\rm op}(\langle \Delta y_n \rangle)$,
which is positive (negative)
when $V_{\rm op}$ is a convex (concave) function.
In the linear regime, however, all corrections $I_m$ ($m\ge 2$) can be ignored.
Moreover it can be shown that the second term on the right-hand side
of Eq.~(\ref{VelocityTotalDerivative}) is also negligible 
in the linear regime (see Appendix~\ref{linear_regime}).
Therefore, the derivation of a macroscopic description
in the linear regime is reduced to developing
a proper approximation of
$\langle \Delta y_n \rangle$.

\subsection{Directed influence}
\label{directed_influence}
A naive approximation of  $\langle\Delta y_n \rangle_{(x,t)}$ 
is $\rho^{-1}(x,t)$. However, this seemingly reasonable approximation
has a serious problem. 
For illustration, it is useful to introduce an unphysical model 
by replacing $\Delta y_n(t)$ in Eq.~(\ref{OptimalVelocityModel}) with $\Delta y_{n-1}(t)$,
so that each vehicle responds to the vehicle {\it behind} it 
rather than the vehicle {\it ahead} of it.
This unphysical model, which differs from the physical one
only by the directionality of the influence, 
has qualitatively different properties.
Thus proper macroscopic descriptions should contain
information about the directionality,
while a naive approximation fails to capture
this information.

To take the directionality into account,
an intuitive prescription was proposed~\cite{Helbing98cond-mat}
without a rigorous justification,
\begin{equation}
\left<\Delta y_n\right>_{(x,t)}\approx \rho^{-1}\left (x+1/2\rho(x,t),t\right ),
\label{midpointapproximation}
\end{equation}
which amounts to evaluating the density at the midpoint between
two vehicles $n$ and $n+1$.
For the above unphysical model, this prescription
results in an expression which is similar to Eq.~(\ref{midpointapproximation})
but has a negative sign in front of $1/2$.
Thus this prescription contains 
information about the directionality.

In the linear regime, we find that a controlled approximation of
$\left<\Delta y_n\right>$ can be obtained
in a rigorous way (see Appendix~\ref{differences}).
The result is
\begin{equation}
\left<\Delta y_n\right> = \rho^{-1}
   +{1\over 2\rho}{\partial \rho^{-1} \over \partial x}
   +\Sigma,
   \label{DistanceExpansion} 
\end{equation}   
where $\Sigma$ represents the sum of all terms 
with second or higher order derivatives.
Note that Eq.~(\ref{DistanceExpansion}) agrees with
the Taylor expansion of the heuristic approximation
[Eq.~(\ref{midpointapproximation})], up to
the first order derivative correction to $\rho^{-1}$.
The deviation occurs in the second order derivative.
While the second order derivative in the Taylor expansion
of Eq.~(\ref{midpointapproximation})
comes with the coefficient $1/8$,
a rigorous calculation leads to the coefficient $1/6$
(see Appendix~\ref{differences}):
\begin{equation}
\Sigma={1\over 6 \rho^2}{\partial^2 \rho^{-1} \over \partial x^2}
+O\left( {\partial^3 \rho^{-1} \over \partial x^3}\right).
\label{SigmaExpansion}
\end{equation}

Thus the leading term in Eq.~(\ref{OptimalVelocityExpansion}) can be expanded as
\begin{equation}
V_{\rm op}\left(\left<\Delta y_n\right>\right) 
= V_{\rm op}(\rho^{-1}) 
+ V'_{\rm op}(\rho^{-1})
\left[{1\over 2\rho}{\partial \rho^{-1} \over \partial x}
   +\Sigma \right]
+ \Sigma_{\rm ir},
   \label{OptimalVelocityFisrtOrderApprox_} 
\end{equation}
where $\Sigma_{\rm ir}$ denotes the sum of linearly irrelevant terms.
By combining Eqs.~(\ref{VelocityTotalDerivative}), (\ref{averageForce}),
(\ref{OptimalVelocityExpansion}), and (\ref{OptimalVelocityFisrtOrderApprox_}),
one obtains
\begin{eqnarray}
   {{\partial v} \over {\partial t}} +
   v{{\partial v}\over{\partial x}}  
 &=&
   \lambda \left[V_{\rm op}\left(\rho^{-1}\right) 
           - v\right] 
  + {\lambda \over{2\rho}}V_{\rm op}^\prime\left(\rho^{-1}\right)
         {{\partial {\rho^{-1}}}\over{\partial x}}
  \nonumber \\
 &+& \lambda V_{\rm op}'\left(\rho^{-1}\right)\Sigma.
   \label{VelocityTotalDeriv_Approx_}  
\end{eqnarray}
Note that the second term proportional to the density gradient
arises from the directed influence, 
while conventional derivations of fluid-dynamic models~\cite{Kerner93PRE}
attribute the density gradient term 
to the velocity variance term in Eq.~(\ref{VelocityTotalDerivative}).
We will call the second term the anticipation term.
The first term is often called the relaxation term.

It is interesting to compare the dispersion relations
of microscopic and macroscopic models.
In a microscopic description,
small perturbations with respect to the homogeneous state
can be written as
\begin{equation}
y_n(t)=v_h t+ \rho_h^{-1}n+\delta y \exp (i\kappa n +\gamma t),
   \label{mic_HomogeneousFlow}
\end{equation}
where $v_h=V_{\rm op}(\rho_h^{-1})$.
By linearizing Eq.~(\ref{OptimalVelocityModel}),
one obtains the dispersion relation
\begin{equation}
\gamma_{\pm} = {\lambda\over2}\left[-1\pm
      \sqrt{1+{4V_{\rm op}^\prime\over\lambda} 
      \left(e^{i\kappa}-1\right)}\,\right].
   \label{mic_Dispersion}
\end{equation}
On the other hand, small perturbations in the macroscopic description
can be written as
\begin{equation}
       \begin{array}{lcl}
         \rho(x,t) &=& \rho_h + \delta \rho \exp(ikx+\omega t),
           \nonumber \\
         v(x,t) &=& v_h + \delta v \exp(ikx+\omega t),
       \end{array}
  \label{mac_HomogeneousFlow}
\end{equation}
where $k\rho_h^{-1}$ is the macroscopic counterpart of $\kappa$
since both represent the phase difference between
two successive vehicles,
and $\omega+ikv_h$ is the macroscopic counterpart of $\gamma$. 
To see the origin of the additional term $ikv_h$, 
note that $\gamma$ is the frequency measured 
in the {\it moving} reference frame with the velocity $v_h$,
while $\omega$ is the frequency measured 
in the stationary frame.
By linearizing Eqs.~(\ref{VehicleConservation}) and (\ref{VelocityTotalDeriv_Approx_}),
one finds
\begin{equation}
\omega_\pm+ikv_h = {\lambda\over2}\left[-1  \pm  
      \sqrt{
       1+{4 V_{\rm op}^\prime\over\lambda} A(k\rho_h^{-1})
      }  \
      \right],
   \label{mac_Dispersion}
\end{equation}
where
\begin{equation}
A(x)= ix+{(ix)^2 \over 2},
\label{A_exp1}
\end{equation}
when the last term in Eq.~(\ref{VelocityTotalDeriv_Approx_}) proportional to $\Sigma$  
is ignored and
\begin{equation}
A(x)= ix+{(ix)^2 \over 2}+{(ix)^3 \over 6},
\label{A_exp2}
\end{equation}
when the leading contribution to $\Sigma$ in Eq.~(\ref{SigmaExpansion})
is included.
Note that $A(x)$ agrees with the Taylor expansion
of the factor $(e^{i\kappa}-1)$ in Eq.~(\ref{mic_Dispersion}).
Thus it is clear that 
the macroscopic momentum equation~(\ref{VelocityTotalDeriv_Approx_}),
combined with the continuity equation~(\ref{VehicleConservation}),
gives a correct description of the long wavelength behavior
of the microscopic model [Eq.~(\ref{OptimalVelocityModel})] 
in the linear regime.

\subsection{Effective diffusion}
\label{effective_diffusion}
Despite the excellent agreement of the long wavelength components, 
it is premature to accept Eq.~(\ref{VelocityTotalDeriv_Approx_})
as a macroscopic momentum equation
since naive treatments of $\Sigma$
introduce an artificial instability,
which is absent in the microscopic model [Eq.~(\ref{OptimalVelocityModel})].
For demonstration, we examine the linear instability criteria.
In the microscopic model, from Eq.~(\ref{mic_Dispersion}) one obtains
that small fluctuations of the mode $\kappa$
become linearly unstable when 
\begin{equation}
V_{\rm op}^\prime \left(\rho_h^{-1}\right) 
> {\lambda \over 1+\cos\kappa}\, .
\label{mic_Criterion}
\end{equation}
Note that the $\kappa=0$ mode shows the strongest instability
and at the critical density where the instability first sets in,
only an infinite wavelength mode becomes unstable. 

In contrast, naive macroscopic models give different results.
When $\Sigma$ is ignored completely, 
Eqs.~(\ref{mac_Dispersion}) and (\ref{A_exp1}) result in
a linear instability criterion $V_{\rm op}'(\rho_h^{-1})>\lambda/2$
for mode $k$. Note that this inequality does not contain $k$.
Thus as soon as $\rho_h$ satisfies this inequality,
fluctuations of {\it all} wavelengths become unstable simultaneously,
different from the behavior in the microscopic description.
On the other hand, when the leading contribution 
to $\Sigma$ in Eq.~(\ref{SigmaExpansion}) is retained,
Eqs.~(\ref{mac_Dispersion}) and (\ref{A_exp2}) result in
$V_{\rm op}'(\rho_h^{-1})>(\lambda/2)[1-(k\rho_h^{-1})^2/6]^{-2}$.
Note that the right-hand side vanishes as $k\rho_h^{-1}\rightarrow \infty$
and thus the homogeneous state is {\it always} unstable
with respect to fluctuations with small wavelengths.
This {\it artificial} instability cannot
be cured by merely using higher order approximations of $\Sigma$.
For example, if we assume that the next order contribution to $\Sigma$
is $(1/4!\rho^3)(\partial^3 \rho^{-1} /\partial x^3)$,
which generates the correct next order in $A(x)$,
one obtains the linear instability criterion
$V_{\rm op}'(\rho_h^{-1})>(\lambda/2)[1-(k\rho_h^{-1})^2/12]/
[1-(k\rho_h^{-1})^2/6]^2$,
which again shows an artificial instability 
for the short wavelength components.

To find the origin of the failure,
it is useful to analyze the microscopic dispersion relation 
[Eq.~(\ref{mic_Dispersion})] 
since the approximations of $\Sigma$ are equivalent to truncating the series
$e^{i\kappa}-1=i\kappa+(i\kappa)^2/2+(i\kappa)^3/3!+(i\kappa)^4/4!+\cdots$
at a certain order.
It can be verified that
when the series is truncated at a {\it finite} order,
highest order terms dominate the physics for large $\kappa$
and generate the artificial instability for large $\kappa$ $(\gg 1)$ modes,
while such instabilities are absent 
when the series is summed up to the {\it infinite} order.
Thus it is clear that truncation at a {\it finite} order
is responsible for the artificial instability.

In this subsection, we aim to develop an approximation of $\Sigma$,
which is compact but still captures
important features of the exact $\Sigma$.
A key observation is that modes 
with $k\rho_h^{-1} \gg 1$ are {\it unphysical}  
since fluctuations on length scales shorter than
the vehicle spacing are not defined
in the original microscopic model.
Motivated by this observation, 
we transform the leading order term of $\Sigma$ in Eq.~(\ref{SigmaExpansion}) 
in such a way that it preserves the same long wavelength behavior
but suppresses fluctuations in short wavelength components 
with $k\gg \rho_h$.
To implement this idea,
one first notes that Eq.~(\ref{VehicleConservation}) relates 
small fluctuations of $\rho$ and $v$ as follows:
\begin{equation}
   \delta \rho = -{{ik\rho_h}\over{\omega + ikv_h}}\, \delta v.
   \label{Fluc_Den_vsVel_}
\end{equation}
One then exploits the correspondence between
$\omega+ikv_h$ and $\gamma$, 
and between $k\rho_h^{-1}$ and $\kappa$.
From the result $\gamma_+\approx 
V_{\rm op}^\prime\left(\rho_h^{-1}\right)i\kappa$ for small $\kappa$,
one obtains 
$$
\delta \rho^{-1} \approx {1\over V_{\rm op}'(\rho_h^{-1})}\,\delta v.
$$
In this derivation, the $\gamma_-$ mode is ignored since
it always decays with time.
Note that the resulting relation amounts to a variational form of
$v = V_{\rm op}(\rho^{-1})$ that can be regarded as the zeroth order 
approximation when $k\rho_h^{-1} \ll 1$.
Its first or higher order corrections will be ignored since 
they introduce third or higher order derivatives to the new approximation 
of $\Sigma$ [Eq.~(\ref{diffusion})].
This way, we construct an approximation
\begin{equation}
   V_{\rm op}^\prime \left(\rho^{-1}\right)\Sigma 
   \approx
         V_{\rm op}^\prime \left(\rho^{-1}\right)
             {1\over 6\rho^2 }
                      {\partial^2 \rho^{-1} \over \partial x^2}
\approx
        {1\over{6\rho^2}}
                      {{\partial^2 v}\over{{\partial x}^2}}\ .
   \label{diffusion}
\end{equation}
The momentum equation becomes
\begin{equation}
    {{\partial v} \over {\partial t}} +
   v{{\partial v}\over{\partial x}} =
   \lambda \left[V_{\rm op}\left(\rho^{-1}\right) 
           - v\right] 
      - {{\lambda V_{\rm op}'} \over{2\rho^3}} 
         {{\partial \rho}\over{\partial x}}
      +  {\lambda \over{6\rho^2}}
                {{\partial^2 v}\over{{\partial x}^2}}\, .
   \label{VelocityTotalDeriv_Final}
\end{equation}
Note that our approximation of $\Sigma$ results in
a {\it diffusion} term, which tends to suppress
short wavelength fluctuations.
Indeed, the linear instability criterion from 
Eqs.~(\ref{VehicleConservation}) and (\ref{VelocityTotalDeriv_Final}) becomes
$V_{\rm op}^\prime \left(\rho_h^{-1}\right) > \lambda(1+k^2/6\rho_h^2)^2/2$,
which confirms the suppression of modes with $k\gg \rho_h$.
In addition, it can be verified that
the macroscopic and microscopic dispersion relations 
agree up to order $k^3$.
Thus we conclude that 
Eq.~(\ref{VelocityTotalDeriv_Final}) is a satisfactory
macroscopic momentum equation in the linear regime. 

Finally, we remark for completeness that
Eq.~(\ref{VelocityTotalDeriv_Final}) {\it cannot} be used to study {\it backward} 
time evolution.
This restriction arises from 
the neglect of the $\gamma_-$ mode,
whose magnitude does {\it grow} in the backward time evolution.

\section{General car-following models}
\label{extension}
In this section, we extend the derivation in Sec.~\ref{optimal}
to general car-following models.
When third or higher order time derivatives
do not appear in microscopic traffic equations,
a general car-following equation with the Galilean invariance
can be written as
\begin{equation}
\ddot y_n=A_{\rm op}(\Delta y_n, \Delta \dot y_n, \dot y_n).
\label{g-mic2}
\end{equation}
Coarse graining leads to
\begin{equation}
  {{\partial v}\over{\partial t}} + v{{\partial v}\over{\partial x}}
    \approx  A_{\rm op} \left(
   \langle \Delta y_n \rangle, \langle \Delta \dot y_n \rangle,v
   \right),
\end{equation}
where $\langle \Delta y_n \rangle$ can be approximated 
by Eqs.~(\ref{DistanceExpansion}) and (\ref{SigmaExpansion}),
and 
\begin{equation}
\langle \Delta \dot y_n \rangle_{(x,t)} \approx  
{1\over \rho}{\partial v\over \partial x}
  +{1\over 2\rho^2}{\partial^2 v\over \partial x^2}\ .
\label{delta_dot_y}
\end{equation}
See Appendix~\ref{differences} for a derivation of
Eq.~(\ref{delta_dot_y}).
We further expand $A_{\rm op}(\cdots)$ as
\begin{eqnarray}
A_{\rm op}(\cdots) &\approx& A_{\rm op}(\rho^{-1},0,v)
 \nonumber \\
&+&A_{\rm op,1}\left({1\over 2\rho}{\partial \rho^{-1} \over \partial x}
  +{1 \over 6\rho^2}{\partial^2 \rho^{-1} \over \partial x^2}\right)
 \nonumber \\
&+&A_{\rm op,2}\left({1\over \rho}{\partial v\over \partial x}
  +{1\over 2\rho^2}{\partial^2 v\over \partial x^2}\right),
  \label{A_expansion}  
\end{eqnarray}
where $A_{{\rm op},i} \equiv \partial_{z_i} A_{\rm op}(z_1,z_2,z_3)|_{(z_1,z_2,z_3)
=(\rho^{-1},0,v)}$.
In real traffic systems,
$A_{\rm op,1}$ and $A_{\rm op,2}$ are expected to be positive
while $A_{\rm op,3}$ is expected to be negative.
Cross-terms proportional to $A_{\rm op,1} A_{\rm op,2}$ are ignored
since they are linearly irrelevant.
The macroscopic momentum equation then becomes
\begin{eqnarray}
{{\partial v} \over {\partial t}} +
   v{{\partial v}\over{\partial x}} 
&=& A_{\rm op}\left(\rho^{-1},0,v \right) 
 +{A_{\rm op,1} \over 2\rho} {\partial \rho^{-1} \over \partial x}
    +{A_{\rm op,1} \over 6 \rho^2} {\partial^2 \rho^{-1} \over \partial x^2}
 \nonumber \\
 &&+  {A_{\rm op,2} \over \rho}{\partial v \over \partial x}
   +{A_{\rm op,2} \over 2 \rho^2} {\partial^2 v \over \partial x^2}.
 \label{gen_eq_of_motion} 
\end{eqnarray}
Note that the dependence of $A_{\rm op}$ on $\Delta \dot y_n$
gives rise to an explicit diffusion term.

Despite the explicit diffusion term,
the artificial instability at short wavelength components
may still arise when $A_{\rm op,1}$ is sufficiently large
since the term proportional to $\partial^2 \rho^{-1}/\partial x^2$
tends to generate the artificial instability,
as demonstrated in Sec.~\ref{optimal}.
Thus we follow the procedure in Sec.~\ref{optimal} B
to obtain
\begin{equation}
{\partial^2 \rho^{-1} \over \partial x^2}\approx
-{A_{\rm op,3} \over A_{\rm op,1}}{\partial^2 v \over \partial x^2}\ ,
\label{short_scale_prescription}
\end{equation}
which is a generalization of Eq.~(\ref{diffusion}).
The resulting momentum equation is
\begin{eqnarray}
{{\partial v} \over {\partial t}} +
   v{{\partial v}\over{\partial x}} 
&=& A_{\rm op}\left(\rho^{-1},0,v \right) 
 +{A_{\rm op,1} \over 2\rho} {\partial \rho^{-1} \over \partial x}
\nonumber \\
 &+&  {A_{\rm op,2} \over \rho}{\partial v \over \partial x}
   +{3A_{\rm op,2}-A_{\rm op,3} \over 6 \rho^2} {\partial^2 v \over \partial x^2}\ .
 \label{gen_eq_of_motion2} 
\end{eqnarray}
Note that the factor $3A_{\rm op,2}-A_{\rm op,3}$ in front of the diffusion term
is manifestly positive.
This equation is free from the artificial instability.

To elucidate the relation with Eq.~(\ref{VelocityTotalDeriv_Final}),
it is useful to define an effective optimal velocity $V_{\rm op,eff}(\rho^{-1})$
in an implicit way as a solution of 
\begin{equation}
A_{\rm op}(\rho^{-1},0,V_{\rm op,eff})=0.
\nonumber
\end{equation}
When $A_{\rm op,3}<0$ for all $v$, the solution is unique
and there is no ambiguity in $V_{\rm op,eff}(\rho^{-1})$.
One also defines 
\begin{equation}
\lambda_{\rm eff}(\rho^{-1},v)\equiv {A_{\rm op}(\rho^{-1},0,v) \over
V_{\rm op,eff}(\rho^{-1})-v}\ ,
\nonumber
\end{equation}
which is positive for all $\rho$ and $v$ if $A_{\rm op,3}<0$ always.
Thus the first term in Eq.~(\ref{gen_eq_of_motion2}) 
can be interpreted as a generalized relaxation term:
\begin{equation}
A_{\rm op}(\rho^{-1},0,v)=\lambda_{\rm eff}(\rho^{-1},v)[V_{\rm op,eff}(\rho^{-1})-v].
\nonumber
\end{equation}

In certain situations, 
the third term in Eq.~(\ref{gen_eq_of_motion2}) 
can be transformed into a familiar form.
One applies the procedure in Sec.~\ref{optimal} B to the term,
and uses the relation
$\gamma_+ \approx -(A_{{\rm op},1}/A_{\rm op,3}) i\kappa (1+\beta i\kappa)$,  
where $\beta \equiv 1/2-A_{\rm op,2}/A_{\rm op,3}-A_{\rm op,1}/A_{\rm op,3}^2$.
Thus we obtain
\begin{equation}
{\partial v \over \partial x} \approx
  -{A_{\rm op,1}\over A_{\rm op,3}} {\partial \rho^{-1} \over \partial x}
 + {\beta \over \rho} {\partial^2 v \over \partial x^2} \, , 
   \label{g-trs2}
\end{equation}
where third or higher order derivatives are neglected.
On the other hand, the second order derivative should be kept
since it renormalizes the diffusion term.
The macroscopic equation of motion then becomes
\begin{equation}
    {{\partial v} \over {\partial t}} +
   v{{\partial v}\over{\partial x}} 
  = \lambda_{\rm eff}[V_{\rm op,eff}-v] 
      - {\nu A_{\rm op,1}  \over 2\rho^3}
       {{\partial \rho}\over{\partial x}}
   - {\mu A_{\rm op,3} \over 6\rho^2}
     {{\partial^2 v}\over{\partial x^2}}\, ,
   \label{DvDt-ef2}
\end{equation}
where $\nu\equiv 1-2A_{\rm op,2}/A_{\rm op,3}$ and 
$\mu\equiv 1-3A_{\rm op,2}/A_{\rm op,3}-6 \beta A_{\rm op,2}/A_{\rm op,3}= 
1-6(A_{\rm op,2}/A_{\rm op,3})(1-A_{\rm op,2}/A_{\rm op,3}-A_{\rm op,1}/A_{\rm op,3}^2)$.
Note that three force density terms in Eq.~(\ref{DvDt-ef2}) 
are in one-to-one correspondence with those 
in Eq.~(\ref{VelocityTotalDeriv_Final}).
Moreover the corresponding terms in the two equations
usually have the same sign 
since $\nu$ is positive and $A_{\rm op,3}$ is negative.
However when $\beta$ in Eq.~(\ref{g-trs2}) is a sufficiently
large negative number, $\mu$ in the diffusion term in Eq.~(\ref{DvDt-ef2})
becomes negative, and an
artificial instability at short wavelength components arises. 
Thus Eq.~(\ref{DvDt-ef2}) can be used
only when $\mu$ is positive
while Eq.~(\ref{gen_eq_of_motion2}) can be used in general situations.

\section{Micro vs Macro}
\label{numerical}
In this section, we compare numerically the properties 
of the microscopic optimal velocity model~[Eq.~(\ref{OptimalVelocityModel})] 
and the macroscopic model
[Eqs.~(\ref{VehicleConservation}) and (\ref{VelocityTotalDeriv_Final})]
derived from it. 
For definiteness, we use
$$V_{\rm op}(\Delta y)={v_{\rm max}\over2}\left[\tanh
\left(2{{\Delta y-x_{\rm neutral}}\over{x_{\rm width}}}
\right)+c_{\rm bias}\right],$$
with $v_{\rm max}=33.6$~m/s, $x_{\rm neutral}=25.0$~m,
$x_{\rm width}=23.3$~m, $c_{\rm bias}=0.913$,
and $\lambda = 2{\rm ~sec}^{-1}$
as in Ref.~\cite{Tadaki98JPSJ}.
A system size $L=2.33$ km is simulated with $N$ vehicles
($\rho_h\equiv N/L$), 
and the following microscopic initial conditions are used:
\begin{equation}
       \begin{array}{ll}
y_n(0)=n\rho^{-1}_h+A\sin(6\pi n\rho^{-1}_h/L), & 1 \le n < N/3,
           \nonumber \\
y_n(0)=n\rho^{-1}_h, & N/3 \le n \le 2N/3,
            \nonumber \\
\dot y_n(0)=V_{\rm op}(\Delta y_n(0)), & \mbox{for all }n. 
       \end{array}
  \label{ini-cdn}
\end{equation}
The corresponding macroscopic initial condition
is prepared by coarse graining the microscopic initial condition
[see Eqs.~(\ref{rhoandQ}) and (\ref{velocity})] 
with the spatial coarse graining function 
$\phi(x,t)=(2\pi\sigma^2)^{-1/2}\mbox{exp}(-x^2/2\sigma^2)\delta(t)$,
where we choose $\sigma=46.4{\rm~m}$.
The periodic boundary condition is imposed
for both the microscopic and macroscopic systems.

We first verify that
the density range $\rho_{\rm c1}<\rho<\rho_{\rm c2}$,
in which the homogeneous traffic state becomes
unstable with respect to infinitesimal perturbations,
is essentially identical 
for the microscopic and macroscopic models.
This implies that, in the linear regime,
the macroscopic model describes
the long wavelength behavior of the microscopic model
very accurately.

To quantify the accuracy of the macroscopic model,
we introduce the space-averaged relative deviation $d_v(t)$,
which is defined by
$$
d_v(t) \equiv { \sqrt{ \langle [v_{\rm macro}(x,t)-v_{\rm micro}(x,t)]^2
\rangle_{\rm space} } \over  \langle v_{\rm micro}(x,t) \rangle_{\rm space} },
$$
where $\langle \cdots \rangle_{\rm space}$ represents the spatial average.
Here $v_{\rm macro}(x,t)$ is calculated from the macroscopic model, while 
$v_{\rm micro}(x,t)$ is obtained by coarse graining
the microscopic configuration at the time $t$.

\begin{figure}
\centerline{\epsfxsize = 8.0cm \epsfbox{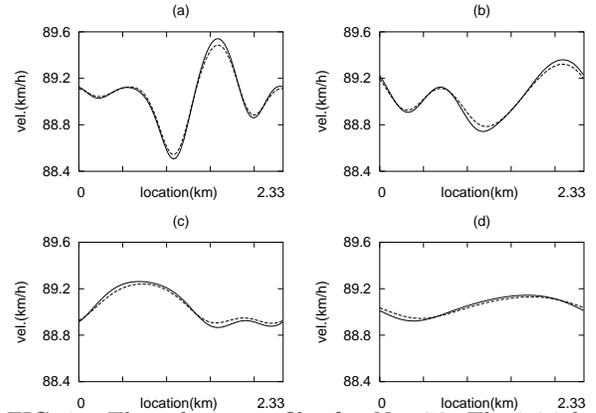}}
\caption{
The velocity profiles for $N=72$. The initial condition in 
Eq.~(\ref{ini-cdn}) is used with $A=1.165$ m.
(a) $t\approx10$~min, (b) $t\approx30$~min, (c) $t\approx1$~h, and
(d) $t\approx4$~h.
The solid (dashed) line shows the microscopic (macroscopic) velocity profile 
in each plot.
The vertical scale is magnified for clarity.
}
\label{stable-profile}
\end{figure}

When the initial perturbation from homogeneous flow is small,
say $A=1.165{\rm~m}$,
we find that $d_v(t)$ is negligible for all density
outside the linearly unstable density range.
A typical velocity profile is shown in Fig.~\ref{stable-profile}.
Note that the macroscopic profiles are almost indistinguishable 
from the microscopic ones.  
Even when $N=72\ (131)$, which corresponds to
a density slightly below (above) the lower (upper) critical
density $\rho_{\rm c1(c2)} \approx 73\ (130)/2.33 {\rm ~km}$ 
(numerically obtained critical densities 
are nearly the same as analytic ones),
$d_v(t)$ remains $\sim 2 \times 10^{-4}$ during
several hours of simulation time.

\begin{figure}
\centerline{\epsfxsize = 8.0cm \epsfbox{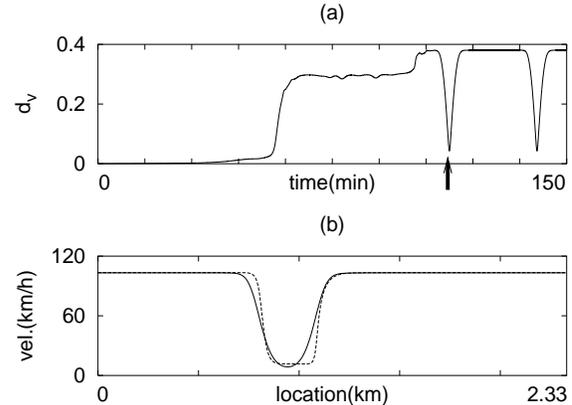}}
\caption{
(a) The time evolution of the space-averaged relative deviation 
of velocity for $A=1.165$ m and $N=73$.
(b) $v_{\rm micro}$ (solid line) vs $v_{\rm macro}$ (dashed line) near 115~min
[marked by the arrow in (a)].
}
\label{ita-fg1}
\end{figure}

The accuracy in the linearly unstable density range is also examined
for $A=1.165{\rm~m}$
and $N=73$, which is the smallest $N$ that demonstrates
the linear instability.
The microscopic simulation shows that the initially smooth profile becomes 
``rough'' as short wavelength fluctuations develop.
An almost identical roughening is found in the macroscopic simulation,
and $d_v(t)$ is almost negligible initially [Fig.~\ref{ita-fg1}(a)].
However, the growth rate of the short wavelength fluctuations is faster in the
microscopic simulation compared to the macroscopic simulation.
This difference is responsible for the rapid growth of $d_v(t)$ near 
$t\approx 55{\rm~min}$. 
The growth of $d_v(t)$ occurs at an earlier time for the density with 
stronger linear instability.
Both in microscopic and macroscopic simulations,
after a sufficient time interval ($\lesssim 120{\rm~min}$) all short
wavelength fluctuations merge into a single large traffic jam,
which moves backward at a constant speed without further evolution in 
its shape. Thus this jam corresponds to the final steady state. 
Figure~\ref{ita-fg1}(b) compares the velocity profiles of the jams from
the microscopic and macroscopic simulations. The velocity of the jam propagation
speed is different and the locations of the jams coincide periodically
in time, resulting in the periodic dips in Fig.~\ref{ita-fg1}(a).

\begin{figure}
\centerline{\epsfxsize = 8.0cm \epsfbox{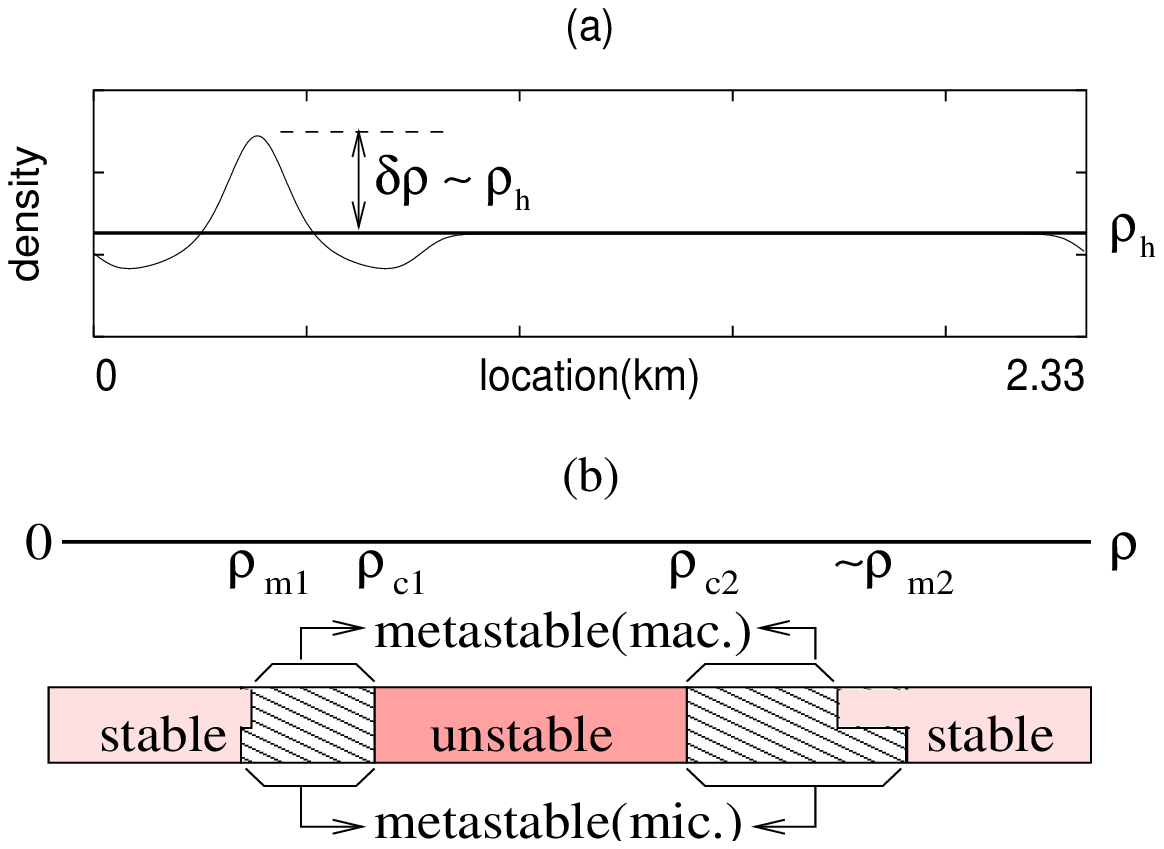}}
\caption{
(a) The density profile for the initial condition in Eq.~(\ref{ini-cdn}).
$\delta \rho$ depends on $A$ and $\rho_h$.
For $A=74.56$ m and $N=100$, $\delta \rho \simeq 1.5 \rho_h$.
(b) Schematic phase diagrams for the microscopic and macroscopic models.
}
\label{dst_rgn}
\end{figure}

Next we choose $A=74.56{\rm~m}$ in Eq.~(\ref{ini-cdn}),
and examine the performance of the macroscopic model
for large perturbations.
Figure \ref{dst_rgn}(a) shows the initial density profile.
After a sufficiently long time, the initial condition
may evolve to a homogeneous state or to a congested state.
The evolution to a congested state is realized 
for $65\lesssim N \lesssim 156$ when the microscopic model is used
and for $66 \lesssim N \lesssim 147$ when the macroscopic model is used.
Thus the lower critical density is in good agreement
while the upper critical density shows about 6\% deviation.
The comparison with the linear critical densities shows that
both microscopic and macroscopic models exhibit {\it metastability},
which implies the hysteresis phenomena in
the metastable density range.
The phase diagram in Fig.~\ref{dst_rgn}(b) summarizes
the result. Note that the microscopic metastable regions are wider.
 
We also investigate the dependence of the critical density 
on $\lambda$ for fixed $A=74.56$ m. 
It is convenient to introduce a dimensionless parameter 
$\bar\lambda\equiv (x_{\rm width}/v_{\rm max})\lambda$, 
which is about $1.387$ for $\lambda=2~{\rm sec}^{-1}$. 
Figure~\ref{vg-ld}(a) shows the relative deviations of the
macroscopic critical densities with respect to the 
microscopic ones. 
For the lower critical density, 
the macroscopic result is in good agreement with
the microscopic one for general $\bar\lambda$.
For the upper critical density, on the other hand,
the deviation of about 6\% at $\bar\lambda \approx 1.387$ shrinks
with the increase of $\bar\lambda$ and
good agreement is achieved near $\bar\lambda=2$.
Thus the difference between the microscopic and macroscopic metastable 
regions in Fig.~\ref{dst_rgn}(b) shrinks
as $\bar\lambda \rightarrow 2$.

The velocity $-v_g$ of a backward propagating traffic jam cluster ($v_g>0$)
is also investigated.
Since $v_g$ is almost independent of $N$,
we fix $N=100$ ($\rho_h\approx 42.9{\rm~km}^{-1}$) for simplicity,
and examine $v_g$ as a function of $\bar\lambda$.
Figure~\ref{vg-ld}(b) (diamonds) shows the ratio
between the microscopic value $v_g^{\rm mic}$ and
the macroscopic value $v_g^{\rm mac}$.
Note that $v_g^{\rm mic}/v_g^{\rm mac}\approx 1$ 
when $\bar\lambda$ is close to 2.
This agreement is notable considering that
the macroscopic model does not have any free parameter 
which can be varied to enhance the agreement.
The agreement, however, becomes less satisfactory
as $\bar\lambda$ becomes smaller.

A crude understanding for the good agreement near $\bar\lambda=2$
can be achieved via the linear analysis, 
although the given initial condition is not in the linear regime.
For the general optimal velocity model,
the linear instability develops when
$\bar V_{\rm op}'>\bar\lambda/(1+\cos \kappa)$;
here we introduce 
$\bar V_{\rm op}' \equiv (x_{\rm width}/v_{\rm max}) V_{\rm op}'$.
This inequality sets an upper limit $\kappa_c$,
above which the instability does not appear.
Note that $\kappa_c$ shrinks to zero as $\bar\lambda/2$ approaches
${\rm max}(\bar V_{\rm op}')$, which is $1$.
Thus the characteristic length scale of the instability 
becomes longer as $\bar\lambda\rightarrow 2$.
This may explain the excellent agreement near $\bar\lambda=2$,
since the macroscopic model becomes more precise
as the characteristic length scale grows.

From these comparisons, we conclude that
the macroscopic model~[Eqs.~(\ref{VehicleConservation}) and (\ref{VelocityTotalDeriv_Final})]
is quite accurate in the linear regime, and 
provides a reasonable description 
of fully developed jam clusters in the nonlinear regime,
although there are deviations in the quantitative level.
But when short length scale dynamics plays an important role,
for example when the avalanchelike growth of many small clusters occurs,
the macroscopic model is not satisfactory.

To construct more accurate macroscopic models, one needs to 
take into account effects of various terms 
ignored in the macroscopic momentum equation derivation. 
As a first trial, we extend the derivation to the nonlinear regime
by including effects of all terms proportional to
$(\partial v/\partial x)^2$, $(\partial \rho^{-1}/\partial x)^2$,
and $(\partial \rho^{-1}/\partial x)(\partial v/\partial x)$
(see Appendix~\ref{nonlinear}).
The resulting equation~(\ref{nonlinear_DvDt_optimal_2}) 
for the same optimal velocity model is examined.
As expected, the linearly unstable density region is identical 
to that by Eq.~(\ref{VelocityTotalDeriv_Final}).
However, the ratio $v_g^{\rm mic}/v_g^{\rm mac}$ deviates further
from one [circles in Fig.~\ref{vg-ld}(b)]. 
Thus it appears that naive inclusion of linearly irrelevant
terms does not improve the accuracy.

\section{Summary}
\label{discussion}
A local macroscopic fluid-dynamic model is derived from
a microscopic car-following model, which establishes
the link between the two types of traffic models.
It is emphasized that the directed influence due
to the breakdown of the balanced action-reaction
is an important ingredient.
For the optimal velocity model, the corresponding
macroscopic momentum equation consists of
a relaxation term, 
an anticipation term (proportional to the density gradient),
and a diffusion term.
Thus it has a structure similar to
the fluid-dynamic model in Ref.~\cite{Kerner93PRE}.
However, the density gradient term is found to arise
from the directed influence rather than the velocity variance.
It is demonstrated that the diffusion term also arises 
from the directed influence.
The derivation provides an unambiguous way to
determine the coefficients of the anticipation term
and the diffusion term. 
The macroscopic model derived from the optimal velocity model
is examined numerically,
and its properties are found to be in reasonable agreement
with those of the microscopic model
although there are deviations in the quantitative level.

\section*{Acknowledgments} 
We acknowledge helpful discussions with H. Y. Lee, who participated 
in the early stage work of this paper.
H. K. L. also thanks G. S. Jeon for discussions.
This work was supported by the Brain Korea 21 Project in 2001.

\appendix
\section{Effects of linearly irrelevant terms}
\label{nonlinear}

While the derivation in Secs.~\ref{optimal} and \ref{extension}
assumes a linear regime,
interesting traffic phenomena often occur
in the nonlinear regime.
In this appendix, we aim to develop a macroscopic momentum equation, 
which is applicable to nonlinear traffic phenomena 
when the characteristic length scale is sufficiently long.
For traffic phenomena with a long characteristic length scale $\xi$, 
each derivative $\partial/\partial x$
can be formally regarded as a small expansion parameter
since it effectively introduces the small factor $1/\xi$.
Then we can take a perturbative approach:
terms without derivatives
constitute the zeroth order contributions,
and terms with the first order derivative 
the first order contributions.
Thus the relaxation and anticipation terms are
the zeroth and first order contributions, respectively.
All zeroth and first order contributions are
already included correctly in Eqs.~(\ref{VelocityTotalDeriv_Final})
and (\ref{gen_eq_of_motion2}).
As for the second order contributions, however,
only part of them are included 
since terms proportional to $(\partial v/\partial x)^2$,
$(\partial \rho^{-1}/\partial x)^2$, or 
$(\partial \rho^{-1}/\partial x)(\partial v/\partial x)$
are of the same order as the diffusion term.
Below we demonstrate a procedure to obtain the missing 
second order contributions
for the general microscopic model [Eq.~(\ref{g-mic2})].

In the general expression~(\ref{VelocityTotalDerivative}), 
the last term proportional to $\partial(\rho \theta)/\partial x$
is irrelevant for our discussion since
it generates third or higher order contributions only
(see Appendix~\ref{linear_regime}).
We expand the first term to obtain
\begin{eqnarray}
\langle \ddot y_n(t')\rangle 
  &\approx& 
  A_{\rm op}\left(\langle \Delta y_n \rangle,\langle \Delta \dot y_n \rangle,
  \dot y_n \right)
  \nonumber \\
&+& {A_{\rm op,11}\over 2}\left\langle \left( \Delta y_n-
  \langle \Delta y_n\rangle\right)^2 \right\rangle
  \nonumber \\ 
&+& {A_{\rm op,22} \over 2} \left\langle \left( \Delta \dot y_n-
  \langle \Delta \dot y_n \rangle \right)^2 \right\rangle
  \nonumber \\
&+& {A_{\rm op,33} \over 2} \left\langle \left( \dot y_n-v \right)^2 \right\rangle
  \nonumber \\
&+& A_{\rm op,12} \langle (\Delta y_n -\langle \Delta y_n \rangle)
(\Delta \dot y_n - \langle \Delta \dot y_n \rangle) \rangle
  \nonumber \\
&+& A_{\rm op,23}\langle (\Delta \dot y_n -\langle \Delta \dot y_n \rangle)
  (\dot y_n -v) \rangle
  \nonumber \\
&+& A_{\rm op,13} \langle (\Delta y_n - \langle \Delta y_n \rangle)
 (\dot y_n -v) \rangle,
  \label{A_nonlinear_exp} 
\end{eqnarray}
which is a generalization of Eqs.~(\ref{averageForce}) and (\ref{OptimalVelocityExpansion}).
Here $A_{{\rm op},ij}\equiv \partial_{z_i}\partial_{z_j}A_{\rm op}(z_1,z_2,z_3)
|_{(z_1,z_2,z_3)=(\rho^{-1},0,v)}$.
In Secs.~\ref{optimal} and \ref{extension}, 
the last six terms in Eq.~(\ref{A_nonlinear_exp}) have been ignored.
For a spatial coarse graining function
$\phi(x,t)=\phi_X(x)\delta(t)$,
we find
\begin{eqnarray}
{A_{\rm op,11}\over 2}\left\langle \left( \Delta y_n-
  \langle \Delta y_n\rangle\right)^2 \right\rangle
  &\approx&{\sigma^2 A_{\rm op,11}\over 2}
  \left( {\partial \rho^{-1} \over \partial x} \right)^2,
  \nonumber \\
{A_{\rm op,33} \over 2} \left\langle \left( \dot y_n-v \right)^2 \right\rangle
  &\approx& {\sigma^2 A_{\rm op,33} \over 2}
  \left({\partial v\over \partial x}\right)^2,
  \nonumber \\
A_{\rm op,13}\langle \left(\Delta y_n  \dot y_n -
  \langle \Delta y_n \rangle v \right) \rangle
  &\approx& \sigma^2 A_{\rm op,13}{\partial v\over \partial x}
  {\partial \rho^{-1} \over \partial x}\ ,
\end{eqnarray}
where $\sigma^2\equiv \int dx' x'^2 \phi_X(x')$.  
Note that these second order contributions depend on 
the coarse-graining function explicitly.
The other three nonlinear terms in Eq.~(\ref{A_nonlinear_exp}) 
give third or higher order contributions only 
(see Appendix~\ref{linear_regime}).

The first term on the right-hand side of Eq.~(\ref{A_nonlinear_exp})
also generates the second order contributions.
The second order expansion of its arguments results in
(see Appendix~\ref{differences})
\begin{eqnarray}
\langle \Delta y_n \rangle &\approx&
  \rho^{-1}+{1\over 2\rho}{\partial \rho^{-1} \over \partial x}
  +{1\over 6\rho^2}{\partial^2 \rho^{-1} \over \partial x^2}
  +{1\over 6\rho}\left( {\partial \rho^{-1} \over \partial x}\right)^2,
  \nonumber \\
\langle \Delta \dot y_n \rangle &\approx&
  {1\over \rho}{\partial v \over \partial x}
  +{1\over 2\rho^2}{\partial^2 v \over \partial x^2}
  +{v\over 2}\left( {\partial \rho^{-1} \over \partial x}\right)^2.
\end{eqnarray}
Thus one finds
\begin{eqnarray}
& & A_{\rm op}(\langle \Delta y_n \rangle,\langle \Delta \dot y_n \rangle,
  \langle \dot y_n \rangle) 
  \nonumber \\
&\approx& A_{\rm op}(\rho^{-1},0,v)
 \nonumber \\
&+& A_{\rm op,1}\left[{1\over 2\rho}{\partial \rho^{-1} \over \partial x}
  +{1 \over 6\rho^2}{\partial^2 \rho^{-1} \over \partial x^2}
  +{1\over 6\rho}\left({\partial \rho^{-1} \over \partial x}\right)^2 \right]
  \nonumber  \\
&+& A_{\rm op,2}\left[{1\over \rho}{\partial v\over \partial x}
  +{1\over 2\rho^2}{\partial^2 v\over \partial x^2}
  +{v \over 2} \left( {\partial \rho^{-1} \over \partial x}\right)^2 \right]
  \\
&+& {A_{\rm op,11} \over 8\rho^2}\left( {\partial \rho^{-1} \over \partial x}
  \right)^2+{A_{\rm op,22}\over 2\rho^2}\left({\partial v\over \partial x}
  \right)^2+{A_{\rm op,12} \over 2\rho^2} {\partial \rho^{-1} \over \partial x}
  {\partial v \over \partial x}.
  \nonumber
\end{eqnarray}
Note that the second order contributions from the expansion of 
$A_{\rm op}(\langle \Delta y_n \rangle,\langle \Delta \dot y_n \rangle,
\langle \dot y_n \rangle)$
do not depend on the coarse-graining function.
Next we apply the prescription
\begin{equation}
{\partial \over \partial x}
\left( {\partial \rho^{-1} \over \partial x}\right)
=-{\partial \over \partial x}
\left({A_{\rm op,3} \over A_{\rm op,1}}{\partial v\over \partial x} \right),
\end{equation}
which is the extension of Eq.~(\ref{short_scale_prescription})
to second order.
The resulting macroscopic momentum equation is
\begin{eqnarray}
&& {\partial v \over \partial t}+v{\partial v \over \partial x}
  \nonumber \\
&=& A_{\rm op}(\rho^{-1},0,v)
  \nonumber \\
&+&{A_{\rm op,1} \over 2\rho}{\partial \rho^{-1} \over \partial x}
+{A_{\rm op,2} \over 2}{\partial v \over \partial x}
+{3A_{\rm op,2}-A_{\rm op,3} \over 6\rho^2}{\partial^2 v \over \partial x^2}
  \nonumber  \\
&+& \left( {A_{\rm op,1} \over 6\rho}+{v A_{\rm op,2} \over 2}
 +{A_{\rm op,11}\over 8\rho^2}+{\sigma^2 A_{\rm op,11} \over 2} \right)
  \left( {\partial \rho^{-1} \over \partial x}\right)^2
  \nonumber  \\
&+& \left( {A_{\rm op,22} \over 2\rho^2}
  -{A_{\rm op,33} \over 6\rho^2}+{A_{\rm op,3}A_{\rm op,13} \over 6\rho^2 A_{\rm op,1}}
  +{\sigma^2 A_{\rm op,33} \over 2}\right)
  \left( {\partial v \over \partial x}\right)^2
  \nonumber \\
&+& \left( {A_{\rm op,12} \over 2\rho^2}
   -{A_{\rm op,13}\over 6\rho^2}
   +{A_{\rm op,3}A_{\rm op,11} \over 6\rho^2 A_{\rm op,1}}
   +\sigma^2 A_{\rm op,13} \right)
   {\partial \rho^{-1} \over \partial x}{\partial v \over \partial x}\ .
  \nonumber \\
  \label{nonlinear_DvDt} 
\end{eqnarray}
For the optimal velocity model [Eq.~(\ref{OptimalVelocityModel})],
this reduces to
\begin{eqnarray}
{\partial v \over \partial t}+v{\partial v \over \partial x}
 & =& \lambda[V_{\rm op}(\rho^{-1})-v]
  + {\lambda V_{\rm op}' \over 2\rho}{\partial \rho^{-1} \over \partial x}
  +{\lambda \over 6\rho^2}{\partial^2 v\over \partial x^2}
  \nonumber \\
&+& \lambda \left( {V_{\rm op}' \over 6\rho}+{V_{\rm op}'' \over 8\rho^2}
  +{\sigma^2 V_{\rm op}'' \over 2} \right)\left( 
  {\partial \rho^{-1} \over \partial x} \right)^2 
  \nonumber \\
&-& {\lambda \over 6\rho^2} {V_{\rm op}'' \over V_{\rm op}'} 
  {\partial \rho^{-1} \over \partial x}
  {\partial v \over \partial x}\ .
  \label{nonlinear_DvDt_optimal} 
\end{eqnarray}

From numerical simulations we find that 
the last two terms give rise to the artificial
instability for short wavelength components
despite the presence of the diffusion term.
It turns out that the artificial instability can be cured
by applying the prescription 
$(\partial \rho^{-1}/\partial x)\approx (1/V'_{\rm op})(\partial v/\partial x)$.
Thus the resulting momentum equation for 
the optimal velocity model reads
\begin{eqnarray}
{\partial v \over \partial t}+v{\partial v \over \partial x}
 & =& \lambda[V_{\rm op}(\rho^{-1})-v]
  + {\lambda V_{\rm op}' \over 2\rho}{\partial \rho^{-1} \over \partial x}
  +{\lambda \over 6\rho^2}{\partial^2 v\over \partial x^2}
  \nonumber \\
&+& {\lambda\over (V'_{\rm op})^2} 
  \left( {V_{\rm op}' \over 6\rho}-{V_{\rm op}'' \over 24\rho^2}
  +{\sigma^2 V_{\rm op}'' \over 2} \right)\left( 
  {\partial v \over \partial x} \right)^2.
  \nonumber \\
  \label{nonlinear_DvDt_optimal_2} 
\end{eqnarray}

\section{Irrelevant terms in the linear regime}
\label{linear_regime}

In this appendix, we assume the spatial coarse graining
$\phi(x,t)=\phi_X(x)\delta(t)$ for definiteness.

(i) $\theta= \langle (\dot y_n-\langle \dot y_n \rangle )^2 \rangle$: 
After some algebra, it can be written as follows,
\begin{eqnarray}
   \theta(x,t) &=&
  {1\over2\rho^2} \int dx^\prime dx^{\prime\prime}  
  \phi_X(x-x^\prime)\phi_X(x-x^{\prime\prime}) 
   \nonumber \\
&\times&\sum_{m,n}
   \left[{\dot y}_m(t)-{\dot y}_n(t)\right]^2
\delta(y_m(t)-x^\prime)
     \delta(y_n(t)-x^{\prime\prime}).
 \nonumber
\end{eqnarray}
When the characteristic length of the variations
is much larger than the coarse-graining scale,
$m-n$ can be formally regarded as small numbers.
To obtain the leading contribution, 
we may then use the formal approximation
$$
\dot y_m(t)-\dot y_n(t)\approx 
\left.{\partial v\over \partial x}\right|_{(x,t)}[y_m(t)-y_n(t)],
$$
which leads to
$$
\theta(x,t)
\approx \left( {\partial v\over \partial x} \right)^2
\left[ \langle y_n^2 \rangle-\langle y_n \rangle^2 \right].
$$
Note that the second factor on the right-hand side
is proportional to the square of the spatial extension
of the coarse-graining function.
When there are many vehicles within the coarse-graining scale,
$$
\langle y_n^2 \rangle-\langle y_n \rangle^2  \approx
\sigma^2,
$$
where $\sigma^2\equiv \int dx' x'^2 \phi_X(x')$.
Thus we obtain
$$
\theta(x,t)\approx \sigma^2
\left( {\partial v\over \partial x} \right)^2.
$$

(ii) $\langle (\Delta y_n-\langle \Delta y_n \rangle )^2 \rangle$:
The procedure is very similar:
\begin{eqnarray}
  &&\left\langle (\Delta y_n -\langle \Delta y_n \rangle )^2 
  \right\rangle_{(x,t)}
  \nonumber \\
  &=&
  {1\over2\rho^2} \int dx^\prime dx^{\prime\prime}  
  \phi_X(x-x^\prime)\phi_X(x-x^{\prime\prime}) 
   \nonumber \\
&\times&\sum_{m,n}
   \left[ \Delta y_m(t)-\Delta y_n(t)\right]^2
\delta(y_m(t)-x^\prime)
     \delta(y_n(t)-x^{\prime\prime}).
 \nonumber
\end{eqnarray}
Using the formal approximation
$$
\Delta y_m(t)-\Delta y_n(t)\approx {\partial \rho^{-1} \over \partial x}
[y_m(t)-y_n(t)],
$$
one finds
$$
\left\langle (\Delta y_n -\langle \Delta y_n \rangle )^2 \right\rangle
\approx \sigma^2 \left( {\partial \rho^{-1} \over \partial x} \right)^2.
$$

(iii) $\langle (\Delta \dot y_n-\langle \Delta \dot y_n \rangle )^2 \rangle$:
\begin{eqnarray}
  &&\left\langle (\Delta \dot y_n -\langle \Delta \dot y_n \rangle )^2 
  \right\rangle_{(x,t)}
  \nonumber \\
  &=&
  {1\over2\rho^2} \int dx^\prime dx^{\prime\prime}  
  \phi_X(x-x^\prime)\phi_X(x-x^{\prime\prime}) 
   \nonumber \\
&\times&\sum_{m,n}
   \left[ \Delta \dot y_m(t)-\Delta \dot y_n(t)\right]^2
\delta(y_m(t)-x^\prime)
     \delta(y_n(t)-x^{\prime\prime}).
 \nonumber
\end{eqnarray}
Since $\langle \Delta \dot y_n \rangle \approx
(1/\rho)(\partial v/\partial x)$ in the leading approximation
(see Appendix.~\protect\ref{differences}),
we use the formal approximation
$$
\Delta \dot y_m(t)-\Delta \dot y_n(t) \approx 
{\partial \over \partial x}
\left( {1\over \rho}{\partial v \over \partial x} \right)
[y_m(t)-y_n(t)],
$$
and obtain
$$
\left\langle (\Delta \dot y_n -\langle \Delta \dot y_n \rangle )^2 
\right\rangle \approx 
\sigma^2 \left[ {\partial \over \partial x} 
\left( {1\over \rho} {\partial v \over \partial x} \right) \right]^2.
$$
 
(iv) $\langle (\Delta y_n- \langle \Delta y_n \rangle)
(\Delta \dot y_n - \langle \Delta \dot y_n \rangle)$:
$$
\langle (\Delta y_n- \langle \Delta y_n \rangle)
(\Delta \dot y_n - \langle \Delta \dot y_n \rangle)
\approx \sigma^2 \ {\partial \rho^{-1} \over \partial x}
{\partial \over \partial x}\left( {1\over \rho} 
{\partial v \over \partial x} \right).
$$
 
(v) $\langle (\Delta \dot y_n -\langle \Delta \dot y_n \rangle )
(\dot y_n -v ) \rangle$:
$$
\langle (\Delta \dot y_n -\langle \Delta \dot y_n \rangle )
(\dot y_n -v ) \rangle
\approx \sigma^2 {\partial \over \partial x}\left(
{1\over \rho}{\partial v \over \partial x} \right)
{\partial v \over \partial x}.
$$

(vi) $\langle ( \Delta y_n- \langle \Delta y_n \rangle)
(\dot y_n -v) \rangle$:
$$
\langle ( \Delta y_n- \langle \Delta y_n \rangle)
(\dot y_n -v) \rangle
\approx \sigma^2 {\partial \rho^{-1} \over \partial x}
{\partial v \over \partial x}.
$$

\section{Macroscopic expressions for the differences}
\label{differences}

This appendix presents derivations of
Eqs.~(\ref{DistanceExpansion}), (\ref{SigmaExpansion}), and (\ref{delta_dot_y}).

(i) $\langle \Delta y_n \rangle$:
One begins with the definition of 
$\langle \Delta y_n\rangle$:
\begin{eqnarray}
\rho\langle \Delta y_n\rangle&=&\int dx' dt' \phi(x-x',t-t')
\nonumber \\
&\times& \sum_n[y_{n+1}(t')-y_n(t')]\delta(y_n(t')-x').
\label{dy_def}
\end{eqnarray}
The following identity is useful:
\begin{equation}
\sum_n \left[y_{n+1}(t)-y_n(t)\right] \delta(y_n(t)-x)
={\partial\over{\partial x}}y_{r(x,t)}(t),
\end{equation}
where $r(x,t)$ is the vehicle number right in front of
$x$ at time $t$. For example, when $y_m(t)<x<y_{m+1}(t)$,
$r(x,t)=m+1$.
Note that each side of
the equation vanishes unless there is a vehicle at $x$,
and that the $x$ integration of each side
from $y_m(t)-\epsilon$ to $y_m(t)+\epsilon$ results
in $y_{m+1}(t)-y_m(t)$, which proves the identity. 
Using the identity, Eq.~(\ref{dy_def}) can be simplified to
\begin{equation}
\langle \Delta y_n\rangle=
\rho^{-1}+\rho^{-1}{\partial \over \partial x}[A_1(x,t)+A_2(x,t)],
\label{dy_exp1}
\end{equation}
where 
\begin{eqnarray}
A_1(x,t) &=& \int dx'dt' \phi(x-x',t-t')
\nonumber \\
&\times& {y_{r(x',t')}(t')-y_{r(x',t')-1}(t') \over 2},
\nonumber \\
A_2(x,t) &=& \int dx'dt' \phi(x-x',t-t')
\nonumber \\
&\times& \left[
{y_{r(x',t')}(t')+y_{r(x',t')-1}(t') \over 2} -x'\right].
\end{eqnarray}
To obtain Eq.~(\ref{dy_exp1}), the integration by parts is used.
Below $\phi(x,t)$ is assumed to be even in $x$.
In the homogeneous state,
$y_{r(x',t')}(t')-y_{r(x',t')-1}(t')=\rho^{-1}$
and $A_1(x,t)=1/2\rho(x,t)$
since $r(x',t')-1$ is the vehicle number right
behind the position $x'$ at time $t'$.
It can also be shown that
$A_2(x,t)=0$ in the homogeneous state.
Thus the first two leading terms in Eq.~(\ref{DistanceExpansion})
can be obtained by replacing $A_1+A_2$ 
in Eq.~(\ref{dy_exp1}) with $1/2\rho$.

To obtain the leading contribution to $\Sigma$ in Eq.~(\ref{SigmaExpansion}),
we calculate $A_1 + A_2 - 1/2\rho$,
which is expected to be proportional to $\partial \rho^{-1} /\partial x$.
However, the term $A_1$ does not give such a contribution.
For illustration, it is useful to
introduce a new coordinate $\tilde x\equiv -x$
and redefine all quantities in terms of the new space
variable. Under this transformation,
$\rho$ and $A_1$ have even parity 
[$\tilde \rho(\tilde x,t)=\rho(- x,t)$,
$\tilde A_1(\tilde x,t)=A_1(-x,t)$],
while the density gradient has the odd parity
[$\partial \tilde \rho^{-1}/\partial \tilde x=-\partial \rho^{-1}/\partial x$].
Since $A_1$ and $\partial \rho^{-1}/\partial x$ have different
parities, $A_1$ should not give a correction
proportional to $\partial \rho^{-1} /\partial x$.

On the other hand, $A_2$ gives a correction proportional to
$\partial \rho^{-1} /\partial x$. 
One uses the identity
\begin{equation}
A_2(x,t)={\partial \over \partial x}\left[
B_1(x,t)-B_2(x,t) \right],
\end{equation}
where
\begin{eqnarray}
B_1(x,t) &=& {1\over 8} \int dx' dt' \phi(x-x',t-t')
  \nonumber \\
 &\times & \left[ y_{r(x',t')}(t')-y_{r(x',t')-1}(t')\right]^2,
  \nonumber \\
B_2(x,t) &=& {1\over 2} \int dx' dt' \phi(x-x',t-t')
  \nonumber \\
  &\times& \left[ { y_{r(x',t')}(t')+y_{r(x',t')-1}(t') \over 2}
  -x'\right]^2.
\end{eqnarray}
In the homogeneous state, $B_1=1/8\rho^2$ and
$B_2=1/24\rho^2$. 
Thus one obtains
\begin{equation}
A_2(x,t)\approx {1\over 12}{\partial \rho^{-2}(x,t)\over \partial x}
={\rho^{-1}(x,t)\over 6}{\partial \rho^{-1}(x,t)\over \partial x}.
\label{A2_result}
\end{equation}
From Eqs.~(\ref{dy_exp1}) and (\ref{A2_result}),
one then finds
$$
\Sigma\approx {1\over 6\rho^2}{\partial^2 \rho^{-1} \over \partial x^2}
+{1\over 6\rho}\left( {\partial \rho^{-1}\over \partial x}\right)^2.
$$

(ii) $\langle \Delta \dot y_n \rangle$:
For derivation, it is convenient
to relate $\langle \Delta \dot y_n \rangle$
to $\langle \Delta y_n \rangle$.
Using integration by parts, one can verify
\begin{equation}
\rho\langle \Delta \dot y_n\rangle =
{\partial \over \partial t}\left[
\rho \langle \Delta y_n \rangle \right]
+{\partial \over \partial x}\left[
\rho \langle \dot y_n \Delta y_n  \rangle \right]. 
\end{equation}
Here $\langle \dot y_n \Delta y_n \rangle$ can be approximated by
$v \langle \Delta y_n \rangle$.
Their difference is proportional to 
$(\partial v/\partial x)(\partial \rho^{-1}/\partial x)$
(see Appendix~\protect\ref{linear_regime}),
and thus we ignore $(\partial /\partial x)[\rho\langle \dot y_n \Delta y_n
\rangle-\rho v \langle \Delta y_n \rangle]$.
One then uses Eq.~(\ref{VehicleConservation}) to obtain
\begin{equation}
\langle \Delta \dot y_n\rangle \approx
\left({\partial \over \partial t}+v{\partial \over \partial x}\right)
\langle \Delta y_n \rangle.
\end{equation}
By using expansion~(\ref{DistanceExpansion}) and 
the continuity equation~(\ref{VehicleConservation}) to convert
temporal derivatives into spatial derivatives,
one obtains 
$$
\langle \Delta \dot y_n \rangle \approx
{1\over \rho}{\partial v \over \partial x}
+{1\over 2\rho^2}{\partial^2 v \over \partial x^2}
+{v\over 2}\left( {\partial \rho^{-1} \over \partial x} \right)^2.
$$ 


\vbox{
\begin{figure}[]
\centerline{\epsfxsize = 8.0cm \epsfbox{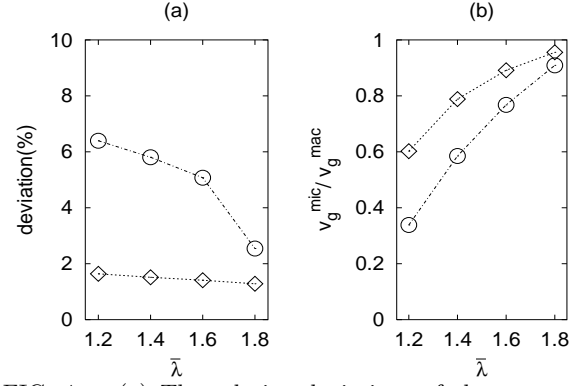}}
\caption {
(a) The relative deviations of the macroscopic lower (diamonds) and 
upper (circles) critical densities with respect to the microscopic 
counterparts for the initial condition [Eq.~(\ref{ini-cdn})] with
$A=74.56$ m. Note that the relative deviations shrink as 
$\bar\lambda$ increases.
(b)The ratio $v_g^{\rm mic}/v_g^{\rm mac}$ 
as a function of $\bar\lambda$
for the macroscopic model [Eqs.~(\ref{VehicleConservation}) and (\ref{VelocityTotalDeriv_Final})] (diamonds)
and for the modified macroscopic model 
[Eqs.~(\ref{VehicleConservation}) and (\ref{nonlinear_DvDt_optimal_2})] (circles)
takes into account the effects of some linearly irrelevant terms.
}
\label{vg-ld}
\end{figure}
}

\end{multicols}
\end{document}